\documentstyle[twoside]{article}

\catcode`\@=11
\long\def\@makefntext#1{
\protect\noindent \hbox to 3.2pt {\hskip-.9pt  
$^{{\eightrm\@thefnmark}}$\hfil}#1\hfill}		%CAN BE USED 

\def\@makefnmark{\hbox to 0pt{$^{\@thefnmark}$\hss}}	%ORIGINAL 
	
\def\ps@myheadings{\let\@mkboth\@gobbletwo
\def\@oddhead{\hbox{}
\rightmark\hfil\eightrm\thepage}   
\def\@oddfoot{}\def\@evenhead{\eightrm\thepage\hfil
\leftmark\hbox{}}\def\@evenfoot{}
\def\sectionmark##1{}\def\subsectionmark##1{}}

\oddsidemargin=\evensidemargin
\newcounter{sectionc}\newcounter{subsectionc}\newcounter{subsubsectionc}
\renewcommand{\section}[1] {\vspace{12pt}\addtocounter{sectionc}{1} 
\setcounter{subsectionc}{0}\setcounter{subsubsectionc}{0}\noindent 
	{\tenbf\thesectionc. #1}\par\vspace{5pt}}
\renewcommand{\subsection}[1] {\vspace{12pt}\addtocounter{subsectionc}{1} 
	\setcounter{subsubsectionc}{0}\noindent 
	{\bf\thesectionc.\thesubsectionc. {\kern1pt \bfit #1}}\par\vspace{5pt}}
\renewcommand{\subsubsection}[1] {\vspace{12pt}\addtocounter{subsubsectionc}{1}
	\noindent{\tenrm\thesectionc.\thesubsectionc.\thesubsubsectionc.
	{\kern1pt \tenit #1}}\par\vspace{5pt}}
\newcommand{\nonumsection}[1] {\vspace{12pt}\noindent{\tenbf #1}
	\par\vspace{5pt}}

\newcounter{appendixc}
\newcounter{subappendixc}[appendixc]
\newcounter{subsubappendixc}[subappendixc]
\renewcommand{\thesubappendixc}{\Alph{appendixc}.\arabic{subappendixc}}
\renewcommand{\thesubsubappendixc}
	{\Alph{appendixc}.\arabic{subappendixc}.\arabic{subsubappendixc}}

\renewcommand{\appendix}[1] {\vspace{12pt}
        \refstepcounter{appendixc}
        \setcounter{figure}{0}
        \setcounter{table}{0}
        \setcounter{lemma}{0}
        \setcounter{theorem}{0}
        \setcounter{corollary}{0}
        \setcounter{definition}{0}
        \setcounter{equation}{0}
        \renewcommand{\thefigure}{\Alph{appendixc}.\arabic{figure}}
        \renewcommand{\thetable}{\Alph{appendixc}.\arabic{table}}
        \renewcommand{\theappendixc}{\Alph{appendixc}}
        \renewcommand{\thelemma}{\Alph{appendixc}.\arabic{lemma}}
        \renewcommand{\thetheorem}{\Alph{appendixc}.\arabic{theorem}}
        \renewcommand{\thedefinition}{\Alph{appendixc}.\arabic{definition}}
        \renewcommand{\thecorollary}{\Alph{appendixc}.\arabic{corollary}}
        \renewcommand{\theequation}{\Alph{appendixc}.\arabic{equation}}
        \noindent{\tenbf Appendix \theappendixc #1}\par\vspace{5pt}}
\newcommand{\subappendix}[1] {\vspace{12pt}
        \refstepcounter{subappendixc}
        \noindent{\bf Appendix \thesubappendixc. {\kern1pt \bfit #1}}
	\par\vspace{5pt}}
\newcommand{\subsubappendix}[1] {\vspace{12pt}
        \refstepcounter{subsubappendixc}
        \noindent{\rm Appendix \thesubsubappendixc. {\kern1pt \tenit #1}}
	\par\vspace{5pt}}

\newcommand{\textlineskip}{\baselineskip=13pt}
\newcommand{\smalllineskip}{\baselineskip=10pt}

\def\eightcirc{
\begin{picture}(0,0)
\put(4.4,1.8){\circle{6.5}}
\end{picture}}
\def\eightcopyright{\eightcirc\kern2.7pt\hbox{\eightrm c}} 

\newcommand{\copyrightheading}[1]
	{\vspace*{-2.5cm}\smalllineskip{\flushleft
	{\footnotesize Modern Physics Letters A #1}\\
	{\footnotesize $\eightcopyright$\, World Scientific Publishing
	 Company}\\
	 }}

\newcommand{\publisher}[2]{{\begin{center}\footnotesize\smalllineskip 
	Received #1\\
	Revised #2
	\end{center}
	}}

\def\abstracts#1#2#3{{
	\centering{\begin{minipage}{4.5in}\footnotesize\baselineskip=10pt
	\parindent=0pt #1\par 
	\parindent=15pt #2\par
	\parindent=15pt #3
	\end{minipage}}\par}} 

\renewenvironment{thebibliography}[1]
	{\frenchspacing
	 \ninerm\baselineskip=11pt
	 \begin{list}{\arabic{enumi}.}
        {\usecounter{enumi}\setlength{\parsep}{0pt}     
	 \setlength{\leftmargin 12.7pt}{\rightmargin 0pt} %FOR 1--9 ITEMS
         \setlength{\itemsep}{0pt} \settowidth
	{\labelwidth}{#1.}\sloppy}}{\end{list}}

\newcounter{itemlistc}
\newcounter{romanlistc}
\newcounter{alphlistc}
\newcounter{arabiclistc}

\newcommand{\fcaption}[1]{
        \refstepcounter{figure}
        \setbox\@tempboxa = \hbox{\footnotesize Fig.~\thefigure. #1}
        \ifdim \wd\@tempboxa > 5in
           {\begin{center}
        \parbox{5in}{\footnotesize\smalllineskip Fig.~\thefigure. #1}
            \end{center}}
        \else
             {\begin{center}
             {\footnotesize Fig.~\thefigure. #1}
              \end{center}}
        \fi}

\newcommand{\tcaption}[1]{
        \refstepcounter{table}
        \setbox\@tempboxa = \hbox{\footnotesize Table~\thetable. #1}
        \ifdim \wd\@tempboxa > 5in
           {\begin{center}
        \parbox{5in}{\footnotesize\smalllineskip Table~\thetable. #1}
            \end{center}}
        \else
             {\begin{center}
             {\footnotesize Table~\thetable. #1}
              \end{center}}
        \fi}

\def\@citex[#1]#2{\if@filesw\immediate\write\@auxout
	{\string\citation{#2}}\fi
\def\@citea{}\@cite{\@for\@citeb:=#2\do
	{\@citea\def\@citea{,}\@ifundefined
	{b@\@citeb}{{\bf ?}\@warning
	{Citation `\@citeb' on page \thepage \space undefined}}
	{\csname b@\@citeb\endcsname}}}{#1}}

\newif\if@cghi
\def\cite{\@cghitrue\@ifnextchar [{\@tempswatrue
	\@citex}{\@tempswafalse\@citex[]}}
\def\citelow{\@cghifalse\@ifnextchar [{\@tempswatrue
	\@citex}{\@tempswafalse\@citex[]}}
\def\@cite#1#2{{$\null^{#1}$\if@tempswa\typeout
	{IJCGA warning: optional citation argument 
	ignored: `#2'} \fi}}
\newcommand{\citeup}{\cite}

\def\pmb#1{\setbox0=\hbox{#1}
	\kern-.025em\copy0\kern-\wd0
	\kern.05em\copy0\kern-\wd0
	\kern-.025em\raise.0433em\box0}

\def\fnt#1#2{\footnotetext{\kern-.3em
	{$^{\mbox{\scriptsize #1}}$}{#2}}}

\def\ps@myheadings{%
    \let\@oddfoot\@empty\let\@evenfoot\@empty
    \def\@evenhead{\slshape\leftmark\hfil}%       %EVEN PAGE
    \def\@oddhead{\hfil{\slshape\rightmark}}%     %ODD PAGE
    \let\@mkboth\@gobbletwo
    \let\sectionmark\@gobble
    \let\subsectionmark\@gobble
    }
\font\tenrm=cmr10
\font\tenit=cmti10 
\font\tenbf=cmbx10
\font\bfit=cmbxti10 at 10pt
\font\ninerm=cmr9

\font\eightrm=cmr8

\def\qed{\hbox{${\vcenter{\vbox{			%HOLLOW SQUARE
   \hrule height 0.4pt\hbox{\vrule width 0.4pt height 6pt
   \kern5pt\vrule width 0.4pt}\hrule height 0.4pt}}}$}}

\newcommand{\ffat}[1]{\mbox {\boldmath $#1$}}

\begin{document}
\setlength{\textheight}{7.7truein}  %for 2nd page onwards

\thispagestyle{empty}

\markboth{\protect{\footnotesize\it The Proton-Deuteron Break-Up process in a Three-Dimensional Approach}}{\protect{\footnotesize\it The Proton-Deuteron Break-Up process in a Three-Dimensional Approach}}

\normalsize\textlineskip

\setcounter{page}{1}

\copyrightheading{}	%{Vol.~0, No.~0 (2002) 000--000}

\vspace*{0.88truein}

\centerline{\bf THE PROTON-DEUTERON BREAK-UP PROCESS}
\baselineskip=13pt
\centerline{\bf IN A THREE-DIMENSIONAL APPROACH}
%\vspace*{0.37truein}
\vspace*{0.4truein}
\centerline{\footnotesize IMAM FACHRUDDIN\footnote{Permanent address: Jurusan Fisika, Universitas Indonesia, Depok 16424, Indonesia}}
\baselineskip=12pt
\centerline{\footnotesize\it Institut f\"ur Theoretische Physik II, Ruhr-Universit\"at Bochum, 44780 Bochum, Germany}
%\baselineskip=10pt
%\centerline{\footnotesize\it City, State ZIP/Zone, Country\footnote{State completely without abbreviations, the affiliation and mailing address, including country. Typeset in 8-pt Times Italic.}}
%\vspace*{10pt}
\vspace*{12pt}

\centerline{\footnotesize CHARLOTTE ELSTER}
\baselineskip=12pt
\centerline{\footnotesize\it Institut f\"ur Kernphysik, Forschungszentrum J\"ulich, 52425 J\"ulich, Germany}
\baselineskip=10pt
\centerline{\footnotesize\it Institute of Nuclear and Particle Physics, Ohio University, Athens, OH 45701, USA}
%\vspace*{0.225truein}
%\vspace*{0.228truein}
\vspace*{12pt}

\centerline{\footnotesize WALTER GL\"OCKLE}
\baselineskip=12pt
\centerline{\footnotesize\it Institut f\"ur Theoretische Physik II, Ruhr-Universit\"at Bochum, 44780 Bochum, Germany}
%\baselineskip=10pt
%\centerline{\footnotesize\it City, State ZIP/Zone, Country}
%\vspace*{0.225truein}
\vspace*{0.228truein}

\publisher{(received date)}{(revised date)}

%\vspace*{0.21truein}
\vspace*{0.23truein}
\abstracts{The pd break-up amplitude in the Faddeev scheme is calculated by employing a three-dimensional method without partial wave decomposition (PWD). In a first step and in view of higher energies only the leading term is evaluated and this for the process d(p,n)pp. A comparison with the results based on PWD reveals discrepancies in the cross section around 200 MeV. This indicates the onset of a limitation of the partial wave scheme. Also, around 200 MeV relativistic effects are clearly visible and the use of relativistic kinematics shifts the cross section peak to where the experimental peak is located. The theoretical peak height, however, is wrong and calls first of all for the inclusion of rescattering terms, which are shown to be important in a nonrelativistic full Faddeev calculation in PWD.}{}{}

%\vspace*{10pt}
%\keywords{The contents of the keywords}

\vspace*{2pt}

%\textlineskip			%) USE THIS MEASUREMENT WHEN THERE IS
%\vspace*{12pt}			%) NO SECTION HEADING

\baselineskip=13pt	        %) ACTUAL LEADING
\normalsize              	%) USE THIS MEASUREMENT WHEN THERE IS
\section{Introduction}		%) A SECTION HEADING
\vspace*{-0.5pt}
\noindent
A three-dimensional (3D) approach without partial wave decomposition (PWD) has been developed for the nucleon-nucleon (NN) system, which uses directly the relative momentum vectors, together with a helicity representation of the total spin. It has been successfully applied to NN scattering\citeup{nn3d} and the deuteron\citeup{deut}, using the realistic NN potentials Bonn-B\citeup{bonnb} and AV18.\citeup{av18} 

We extend this approach to the three-nucleon (3N) break-up process. In a first step we consider only the leading term in the multiple scattering series, which is in first order in the NN t-matrix $T$. The interest is twofold: we want to see the onset of the limitations of a standard PWD by going to higher energies and whether this first order treatment is sufficient. In addition, the shifts in the observables by changing from nonrelativistic to relativistic kinematics will be investigated.

\newpage
\section{Formulation}
\noindent
The leading term of the pd break-up amplitude in the Faddeev scheme is
given by
\begin{equation}
U_{0} = \left\langle {\bf p}{\bf q}m_{1}m_{2}m_{3}\tau _{1}\tau _{2}\tau _{3}\right| (1+P)TP\left| \phi \right\rangle = U_{0}^{(1)}+U_{0}^{(2)}+U_{0}^{(3)}\, ,
\end{equation}
where ${\bf p}, \, {\bf q}$ are Jacobi momenta, $m_{i,}\tau _{i}\, (i=1,2,3) $ spin and isospin quantum numbers, $T$ the NN t-matrix, $P=P_{12}P_{23}+P_{13}P_{23}$ permutation operators, $\left| \phi \right\rangle = \left| {\bf q}_{0}m^{0}_{1}\tau ^{0}_{1}\right\rangle \left| \varphi _{d}M_{d}\tau _{d}\right\rangle $ the initial state and $U_0^{(i)}\, (i=1,2,3) $ the three parts of the amplitude resulting from $(1+P)$. 

Now $U_0$ can be expressed in terms of the t-matrix $T^{\pi St}_{\Lambda \Lambda '}(q,q',\cos \theta ,E)$ in the momentum-helicity basis, for given parity, total spin and isospin, and final and initial
helicities $\Lambda '$ and $ \Lambda$.\citeup{nn3d} Since $U_0^{(2)},\, U_0^{(3)}$ are related to $U_0^{(1)}$ by means of permutations, it is sufficient to work out only $U_0^{(1)}$. We obtain
\begin{eqnarray}
U_{0}^{(1)}({\bf p},{\bf q}) & = & (-)^{\frac{1}{2}+\tau _{1}}\delta _{\tau _{2}+\tau _{3},\tau ^{0}_{1}-\tau _{1}}\frac{1}{4\sqrt{2}}\sum _{m'}\sum _{l=0,2}Y_{l,M_{d}-m'-m_{1}}(\hat{\pi }')\varphi _{l}(\pi ')\nonumber\\
 &  & C\left( \frac{1}{2}\frac{1}{2}1;m_{1}m',m'+m_{1}\right) C\left( l11;M_{d}-m_{1}-m',m_{1}+m'\right) \nonumber\\
 &  & \sum _{S\pi t}\left( 1-\eta _{\pi }(-)^{S+t}\right) C\left( \frac{1}{2}\frac{1}{2}t;\tau _{2}\tau _{3}\right) C\left( \frac{1}{2}\frac{1}{2}t;\tau _{1}^{0},-\tau _{1}\right) e^{-i(\Lambda _{0}\phi -\Lambda _{0}'\phi _{\pi })} \nonumber\\
 &  & C\left( \frac{1}{2}\frac{1}{2}S;m_{2}m_{3}\Lambda _{0}\right) C\left( \frac{1}{2}\frac{1}{2}S;m_{1}^{0}m'\Lambda _{0}'\right) \sum _{\Lambda \Lambda '}d^{S}_{\Lambda _{0}\Lambda }(\theta )d^{S}_{\Lambda _{0}'\Lambda '}(\theta _{\pi }) \nonumber\\
 &  & e^{i(\Lambda '\phi ''-\Lambda \Omega )}T_{\Lambda \Lambda '}^{\pi St}\left( p,\pi ,\cos \theta '';E_{d}+\frac{3}{4m}\left( q_{0}^{2}-q^{2}\right) \right) 
\end{eqnarray}
with ${\ffat \pi }\equiv \frac{1}{2}{\bf q}+{\bf q}_{0}$ and ${\ffat \pi }'\equiv -{\bf q}-\frac{1}{2}{\bf q}_{0}$. Besides standard notations there occur the deuteron wave function components $\varphi _{l}(\pi ')$. $ U_{0}^{(2)} $ and $ U_{0}^{(3)} $ are obtained by suitable replacements of the momenta and discrete quantum numbers. 

\section{Results and Discussions}
\begin{figure}[!h]
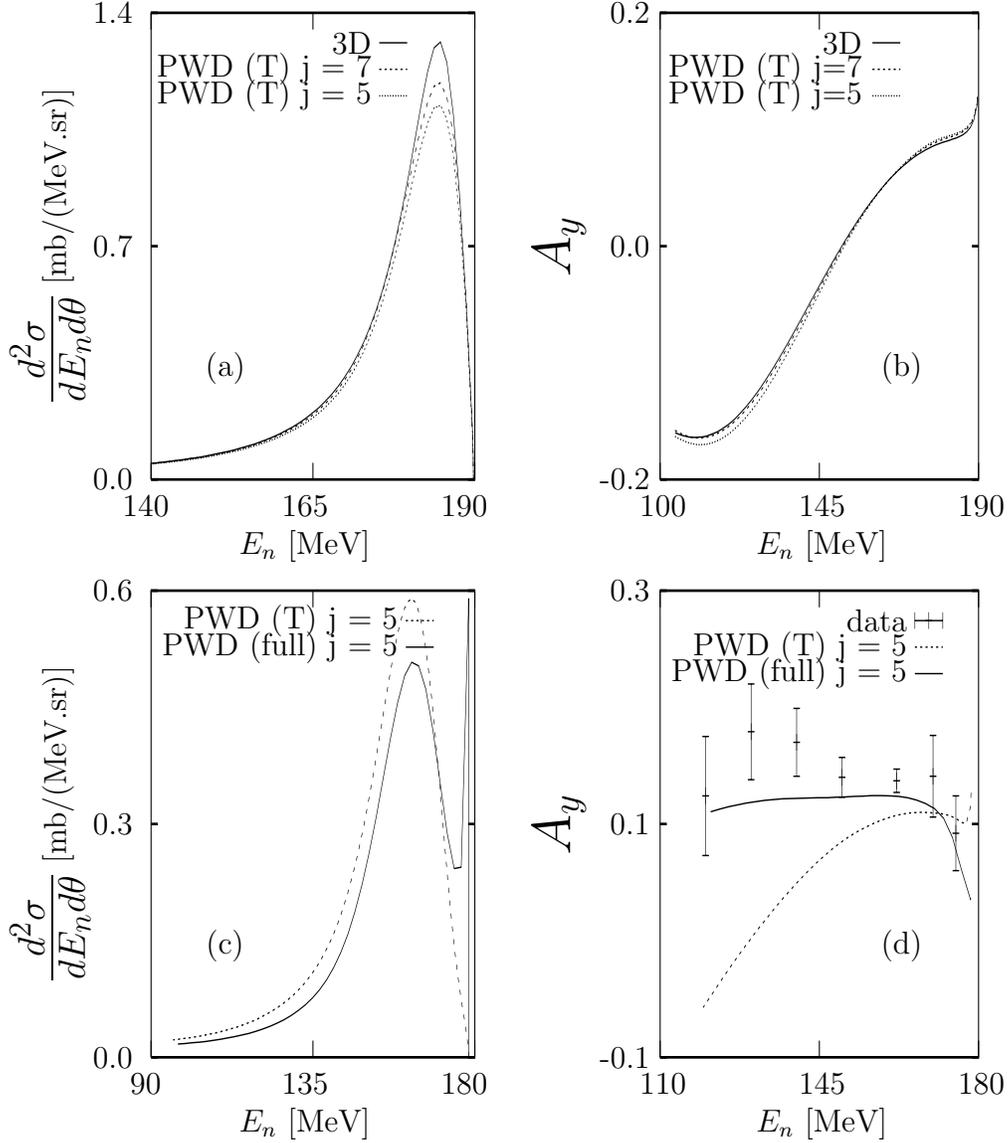

\begin{minipage}[t]{60mm}
\begin{center}
\input{d213spec.197.tex}
\end{center}
\end{minipage}
\hfill
\begin{minipage}[t]{60mm}
\begin{center}
\input{ay13spec.197.tex}
\end{center}
\end{minipage}
\newline
\begin{minipage}[t]{60mm}
\begin{center}
\input{d224full.197.tex}
\end{center}
\end{minipage}
\hfill
\begin{minipage}[t]{60mm}
\begin{center}
\input{ay24full.197.tex}
\end{center}
\end{minipage}
\caption{\label{fig1}The spin averaged differential cross section (a \& c) and the analyzing power $A_y$ (b \& d) at $E_{lab}=197$ MeV, $\theta = 13^0$ (a \& b) and $\theta = 24^0$ (c \& d) for the process $d(p,n)pp$.}
\end{figure}

\noindent
Here we use the NN potentials Bonn-B and AV18. Figs.\ref{fig1}(a) and (b) compare our 3D results to the ones based on the PWD, both first oder in $T$, for the cross section and the analysing power $A_y$ in the process $d(p,n)pp$. The PWD calculation includes the NN t-matrix for total NN angular momenta $j=5$ and $7$, and states of total 3N angular momenta up to $J=31/2$. We see a significant discrepancy in the peak of the cross section but agreement for $A_y$ in case of $j = 7$. Since $j=7$ is not feasible right now in a full Faddeev calculation (all orders in $T$) one has apparently reached the limits of a generally reliable PWD at such an energy of about 200 MeV. At lower energies, say 100 MeV, the PWD agrees perfectly well with our 3D results. In Figs.\ref{fig1}(c) and (d) we
show the full Faddeev results for the cross section and $A_y$, now  based on $j=5$ in the PWD. The figurues reveal that rescattering is quite important, even at this relatively high energy.

\begin{figure}[!h]
\begin{center}
\input{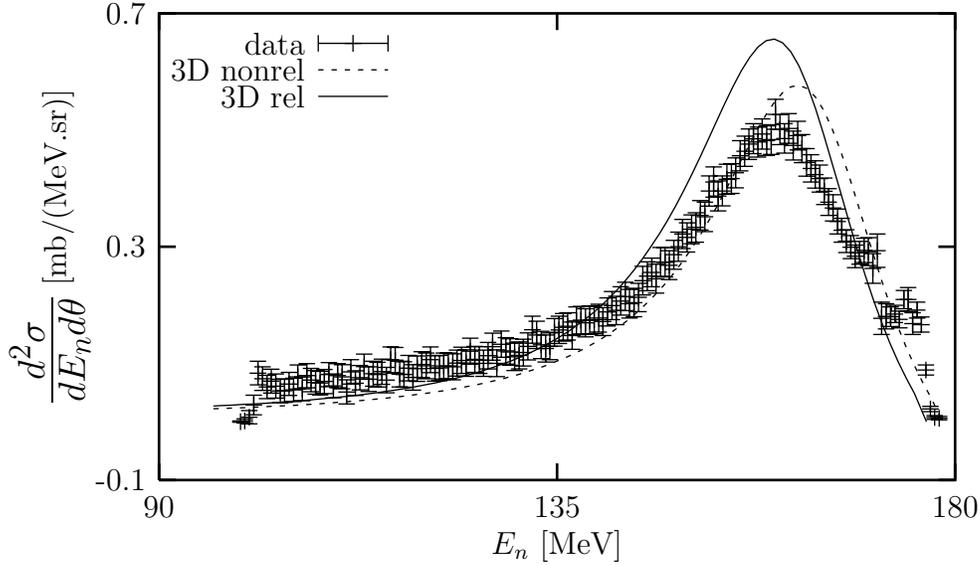}
\end{center}
\caption{\label{fig2}The spin averaged differential cross section for the process $d(p,n)pp$ at $E_{lab}=197$ MeV, $\theta = 24^0$.}
\end{figure}

In Fig.\ref{fig2} we compare our calculated cross section to data\citeup{data197}. The 3D result (called ``3D nonrel'') clearly deviates by giving the peak at the wrong position. Next we use relativistic kinematics by employing appropriately Lorentz boosted momenta\citeup{rel}. This shifts the curve ``3D nonrel'' to the one labeled ``3D rel'', which gives the correct peak position. 
However, this calculation overshoots the data. Thus, our
calculations of the break-up cross section indicate that one
needs altogether three ingredients: a 3D treatment, rescattering effects and relativity. For $A_y$ shown in Fig.\ref{fig1}(d) the full nonrelativistic Faddeev calculation in the PWD is closed to the data. Since the studied relativistic effects for $A_y$ are quite small (not shown) and the higher partial waves appear unimportant as also shown in Fig.\ref{fig1}(b) that agreement will presumably survive. In case of the cross section we see in Fig.\ref{fig1}(c) a decrease due to rescattering, which might cure the overshooting seen in the relativistic curve of first order in $T$ in Fig.\ref{fig2}. More details, especially on various spin observables in the process $d(\vec{p}, \vec{n})pp$ at 197 MeV and higher energies can be found in\citeup{thesis} and in  a forthcoming publication.

\nonumsection{Acknowledgments}
\noindent
We would like to thank H. Witala and J. Golak from Instytut Fizyki, Uniwersytet Jagiellonski, Cracow, Poland for providing the PWD calculations.

\nonumsection{References}


\begin{thebibliography}{000}
\bibitem{nn3d}
I. Fachruddin, Ch. Elster, W. Gl\"ockle, {\it Phys. Rev.} {\bf C62}, 044002 (2000).

\bibitem{deut}
I. Fachruddin, Ch. Elster, W. Gl\"ockle, {\it Phys. Rev.} {\bf C63}, 054003 (2001).

\bibitem{bonnb}
R. Machleidt, {\it Adv. Nucl. Phys.} {\bf 19}, 189 (1989)

\bibitem{av18}
R. B. Wiringa, V. G. J. Stoks, and R. Schiavilla, {\it Phys. Rev.} {\bf C51}, 38 (1995).

\bibitem{data197}
D. L. Prout, et.al., {\it Phys. Rev.} {\bf C65}, 034611 (2002).

\bibitem{rel}
R. Fong and J. Sucher, {\it J. Math. Phys.} {\bf 5}, 456 (1964).

\bibitem{thesis}
I. Fachruddin, PhD. thesis, Ruhr-Universit\"at Bochum, 2002.

\end{thebibliography}
\end{document}